\documentclass[letterpaper, 10 pt, conference]{ieeeconf}  % Comment this line out if you need a4paper

\IEEEoverridecommandlockouts                              % This command is only needed if 
\usepackage{graphicx} % for pdf, bitmapped graphics files
\usepackage{amsmath} % assumes amsmath package installed
\usepackage{amssymb}  % assumes amsmath package installed
\usepackage{cite}

\title{\LARGE \bf
Time-varying System Identification of Bedform Dynamics Using Modal Decomposition}

\author{Shakib Mustavee (Member IEEE)$^{1}$, Arvind Singh, and Shaurya Agarwal (Senior Member IEEE)$^{2}$% <-this % stops a space
\thanks{*A. Singh acknowledges partial support from the U.S. National Science Foundation under Grant EAR-2342936.}% <-this % stops a space
\thanks{$^{1}$Shakib Mustavee is a Post-doc with the Department of Civil, Environmental \& Construction Engineering, University of Central Florida, Orlando, Florida.
        {\tt\small shakib.mustavee@ucf.edu}}%
\thanks{$^{2}$ Arvind Singh is an Associate Professor with the Department of Civil, Environmental \& Construction Engineering, University of Central Florida, Orlando, Florida.
        {\tt\small arvind.singh@ucf.edu}}%        
\thanks{$^{2}$ Shaurya Agarwal is an Associate Professor with the Department of Civil, Environmental \& Construction Engineering, University of Central Florida, Orlando, Florida.
        {\tt\small shaurya.agarwal@ucf.edu}}%
}

\begin{document}

\maketitle
\thispagestyle{empty}
\pagestyle{empty}

%%%%%%%%%%%%%%%%%%%%%%%%%%%%%%%%%%%%%%%%%%%%%%%%%%%%%%%%%%%%%%%%%%%%%%%%%%%%%%%%
\begin{abstract}
Measuring sediment transport in riverbeds has long been
a challenging research problem in geomorphology and river
engineering. Traditional approaches rely on direct measurements using sediment samplers. Although such measurements are often considered ground truth, they are intrusive, labor-intensive, and prone to large variability. As an alternative, sediment flux can be inferred indirectly from the kinematics of migrating bedforms and temporal changes in bathymetry. While such approaches are helpful, bedform dynamics are nonlinear and multiscale, making it difficult to determine the contributions of different scales to the overall sediment flux. Fourier decomposition has been applied to examine bedform scaling, but it treats spatial and temporal
variability separately. In this work, we introduce Dynamic
Mode Decomposition (DMD) as a data-driven framework for
analyzing riverbed evolution. By incorporating this
representation into the Exner equation, we establish a link
between modal dynamics and net sediment flux. This
formulation provides a surrogate measure for scale-dependent
sediment transport, enabling new insights into multiscale
bedform-driven sediment flux in fluvial channels.
\end{abstract}

\section{Introduction}\label{sec:intro}

Accurately modeling and predicting sediment transport in fluvial environments remains a fundamental challenge in geomorphology and river engineering. Fluvial channels act as primary conduits for both water and sediment \cite{lee2023scaling,lee2022reconstructing}. The continuous processes of erosion and deposition shape Earth's surface, influence ecological diversity, and affect human activities such as navigation, flood control, and infrastructure development \cite{wang2016reduced}. Since sediment transport underpins both the morphology and function of channel systems, reliable approaches for its quantification are of critical importance. 

Traditionally, sediment transport has been measured directly using devices such as sediment traps or Helley–Smith samplers, which capture material moving along the bed over a specified interval \cite{vericat2006bed,helley1971development,singh2009experimental,bunte2005effect}. While often regarded as ground truth, these methods are labor-intensive, intrusive, and highly sensitive to deployment conditions. Reported measurements vary substantially in space and time, in addition to factors such as sampler orientation, mesh size, sediment grain distribution, and the hydrodynamic disturbance created by the device itself. As a result, direct sampling is not only time-consuming but also limited in accuracy and duration, making it insufficient for resolving the dynamics of sediment flux in natural channels. Given these limitations, researchers have increasingly turned to indirect approaches. One widely adopted method is to infer sediment flux from the kinematics of migrating bedforms and from temporal changes in channel bathymetry (see for details \cite{singh2012coupled,lee2023scaling,singh2011multiscale}). Bedforms act as signatures of sediment motion.  
Their characteristics, such as amplitude, wavelength, and migration velocity, can provide indirect estimates of volumetric sediment flux \cite{simons1965bedload}. The volumetric sediment flux, $q_{s}$, represents the rate at which sediment volume passes per unit channel width.  
Estimating $q_{s}$ is central to quantifying bedload transport. However, bedform dynamics are inherently nonlinear and strongly scale dependent \cite{ranjbar2020entropy}. Bedforms of different sizes coexist, migrate at distinct velocities, and interact through feedbacks among morphology, shear stress, and transport rate \cite{lee2023scaling,singh2011multiscale,guala2020mixed}. This multiscale variability makes it difficult to determine the relative contributions of different bedform scales to the overall sediment flux, which remains an open research question. Previous studies have used Fourier decomposition of bed elevation fields to examine scaling behavior and the role of different bedform sizes in sediment transport \cite{guala2014spectral}. These analyses showed that small, fast-migrating secondary bedforms can significantly contribute to flux and drive the propagation of larger, slower bedforms. While Fourier methods provide valuable spectral information, they treat spatial and temporal variations separately.

\begin{figure}[t]
    \centering
    \includegraphics[width=1.0\linewidth]{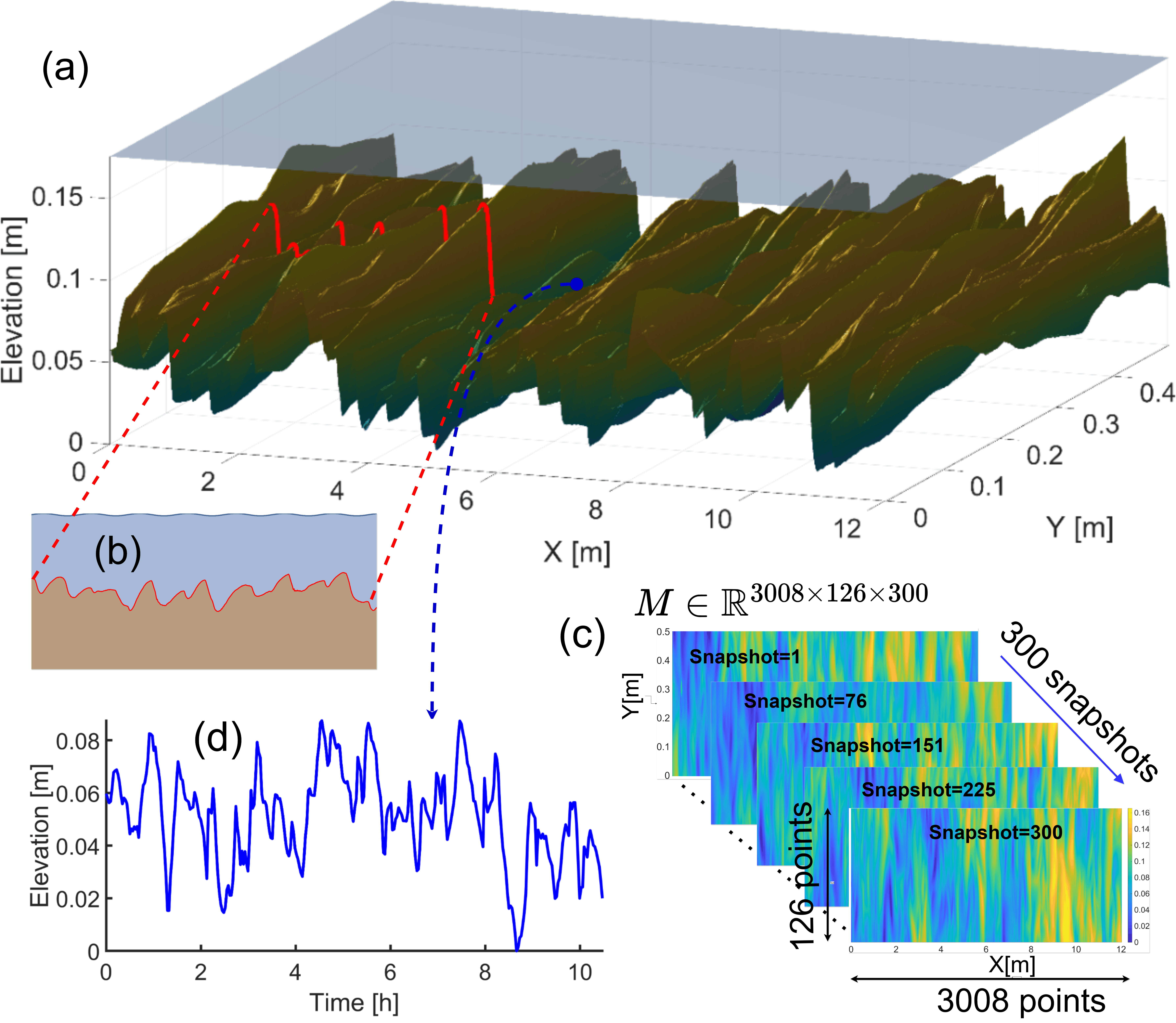}
    \caption{Riverbed topography visualization. (a) Illustrates 2D topography at a specific time. (b) Spatial bed elevation profile along a selected transect (shown in red line in Figure 1(a)). (c) Spatio-temporal evolution of the riverbed topography. The figure also describes the dimensions of the data. (d) Temporal bed elevation profile along a selected point.}
    \label{fig:schematic}
\end{figure}

Dynamic Mode Decomposition (DMD) has emerged as a widely used tool in fluid mechanics and data-driven modeling, valued for its ability to extract coherent spatio-temporal patterns and provide reduced-order representations of complex flow fields \cite{higham2018implications,sharma2016correspondence,mcelroy2009nature}. In this work, we propose a novel framework based on DMD for characterizing the scale dependence of sediment transport analysis. DMD decomposes the evolution of bed elevation into spatio-temporal modes, each representing a coherent pattern with an associated growth/decay rate and frequency. This makes DMD particularly well-suited for the ``tall-and-skinny" structure of riverbed elevation datasets, where many spatial samples are collected over relatively short time records \cite{kutz2016dynamic}. By substituting the DMD-based linear approximation of bed elevation dynamics into the Exner equation and integrating across space, we establish a connection between DMD modes and net sediment flux. 

%, andreuzzi2023dynamic,

%This formulation provides a surrogate measure of sediment transport that explicitly accounts for the contribution of different scales, offering new insights into the multiscale dynamics of bedform-driven sediment flux in fluvial channels.

\noindent \textbf{Contributions:} The contributions of this article are as follows:

\begin{itemize}
    \item The paper studies the spatio-temporal coherent structure of the riverbed elevation using DMD. 
    
    \item It derives a mathematical connection between sediment flux and spatio-temporal modes of riverbed elevation computed from DMD.
\end{itemize}

\section{Data Description}\label{section:data}
The dataset analyzed in this study originates from large-scale flume experiments conducted at the St. Anthony Falls Laboratory (SAFL) Main Channel. The flume is \(85\,m\) long and horizontally \(2.75\,m\) wide and is equipped with discharge regulation and sediment recirculation systems to maintain morphodynamic equilibrium during experiments \cite{lee2022reconstructing}. The spatio-temporal bed evolution was measured along the channel centerline using a submerged laser scanning device mounted on an automated cart. The dataset consists of three-dimensional arrays of bed elevation change, denoted as $\eta(x,y,t)$, where $t$ represents the acquisition time in seconds, $x$ the streamwise locations, and $y$ the spanwise locations. Formally,
\(
\eta(x,y,t) \in \mathbb{R}^{N_x \times N_y \times N_t},
\)
where $N_x$, $N_y$, and $N_t$ correspond to the number of streamwise, spanwise, and temporal indices, respectively. Figure~\ref{fig:schematic} explains the bed elevation data. Figure~\ref{fig:schematic} (a) illustrates the topography at a given moment. In the stream direction, the bed is \(12 \;m\) long and \(0.5 \; m\) wide. The elevation of the bed is limited between \(0\) to \(0.329 \;m\). There are \(3008\) points in the stream direction \((N_{x}=3008)\) and there are \(126\) points in the span direction (\(N_{y} = 126\)).  Figure~\ref{fig:schematic} (b) shows the elevation of the river bed along a transect. Figure~\ref{fig:schematic}(c) shows the temporal evolution of the topography by representing the three-dimensional bed elevation in Figure~\ref{fig:schematic}(a) as a two-dimensional heat map, which also illustrates the dimensions of the spatio-temporal data. Figure~\ref{fig:schematic} (d) shows the temporal evolution of a point at the middle of the channel on a transect. The data was recorded for \(9.71\) hours at a resolution of approximately \(2 \; \text{min}\), which corresponds to \(300\) time snapshots. This high-resolution spatio-temporal record of channel bed morphology provides an ideal basis for applying dynamic mode decomposition (DMD) to investigate the coherent structures and temporal dynamics of riverbed evolution. 

\section{DMD Framework for Bed Elevation} \label{sec:math}
% DMD on steady flume data reveals dominant modes clustered near the unit circle, consistent with persistent bedforms, and allows mode-wise transport ranking. Under synthetic transients, certain eigenvalues migrate inside/outside the unit circle, indicating decaying or growing features. Mode-wise transport decomposition highlights which modes dominate steady transport and how contributions shift after a forcing event. A brief sensitivity test shows that downsampling (up to 10× reduction) preserves the identity of dominant modes, suggesting robustness in sensing requirements.
\subsection{Formulation of DMD for Bed-Elevation Fields}
DMD is a data-driven technique that extracts coherent
spatio-temporal patterns from data by decomposing it into
modes and complex eigenvalues. Each mode represents a spatial
structure, while each eigenvalue characterizes the temporal
behavior through a growth/decay rate and oscillation
frequency \cite{kutz2016dynamic}. In practice, DMD assumes
that consecutive snapshots of the system satisfy an
approximate linear evolution of the form $X' = AX$. Here,
$X=[x_1,x_2,\ldots,x_{m-1}]$ and
$X'=[x_2,x_3,\ldots,x_m]$ are snapshot matrices constructed
from sequential measurements of the system state, where each
column $x_k$ represents a vectorized snapshot of the
spatial field at time $t_k$. The operator $A$ is an unknown
linear mapping that advances the state from one snapshot to
the next. DMD estimates this operator directly from data and
computes its eigenvalues and eigenvectors to obtain the
dominant dynamical modes. This procedure yields a
reduced-order linear representation of the system dynamics,
which can often be interpreted as a finite-dimensional
approximation of the Koopman operator
\cite{mustavee2025koopman,mustavee2022linear}.

We apply DMD to analyze the temporal evolution of riverbed topography described in Section~\ref{section:data}.
First, we consider a sequence of bed-elevation fields $\{\eta_1(x,y),\; \eta_2(x,y),\; \dots,\; \eta_p(x,y)\}$ measured on a Cartesian grid with \(m\) rows and \(n\) columns measured at discrete times \(t_1,\dots,t_p\). Each snapshot can be written as: 
\[
\eta_k \equiv \eta_k(x,y) \in \mathbb{R}^{m\times n}, \qquad \forall k=1,\dots,p.
\]

\begin{figure}
    \centering
    \includegraphics[width=1.0\linewidth]{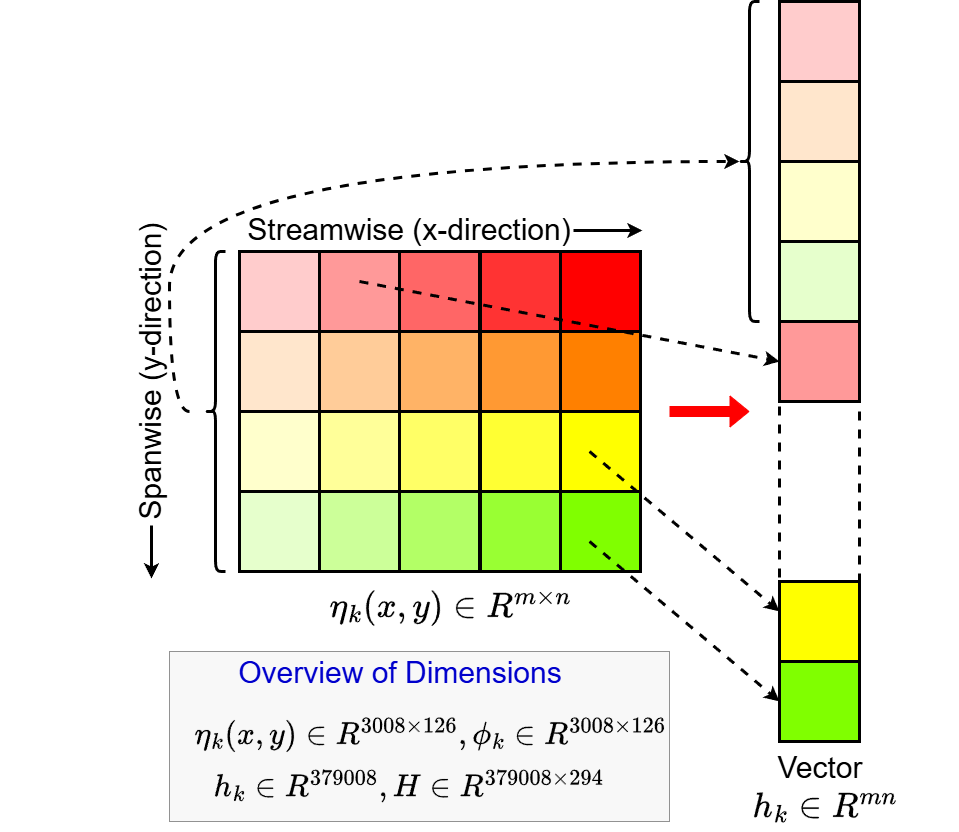}
    \caption{Schematic showing how a two-dimensional bed elevation snapshot is stacked into a one-dimensional column vector for constructing the DMD data matrix.}
    \label{fig:vectomat}
\end{figure}

Figure~\ref{fig:vectomat} shows how the two-dimensional snapshot is vectorized by stacking each row using the \(\operatorname{vec}(\cdot)\) operator:
\[
h_k \;=\; \operatorname{vec}\big(\eta_k\big) \in \mathbb{R}^{mn}, \qquad k=1,\dots,p.
\]
By arranging the vectorized snapshots into a single matrix, we get the spatio-temporal matrix $H$, where, $H\;=\; \big[\,h_1,\; h_2,\; \dots,\; h_p\,\big] \in \mathbb{R}^{\,mn\times p}.$ Here, \(m\times n\) denotes the size of the spatio-temporal matrix while \(mn\) is the length of each vectorized snapshot, and \(p\) denotes the number of snapshots. We create two metrices, \(H_{1}\) and its time shifted version \(H_{2}\) from \(H\) by choosing the first \(q+1\) number of snapshots: 
\(
H_1 \;=\; \big[\,h_1,\; h_2,\; \dots,\; h_{q}\,\big] \; \text{and} \;
H_2 \;=\; \big[\,h_2,\; h_3,\; \dots,\; h_{q+1}\,\big]
\). where, \(H_{1},H_{2}\in\mathbb{R}^{\,mn\times q}\). We assume a linear map \(A\) such that
\begin{equation}\label{eq:Ah}
H_2 \approx A\,H_1 \implies A \approx H_{2}H^{\dagger}_{1}
\end{equation}
Here, \(A\in\mathbb{R}^{\,mn\times mn}\) and \(H^{\dagger}_{1}\) is the Moore-Penrose pseudo-inverse of \(H_{1}\). Directly solving for \(A\) is computationally expensive due to the high-dimensionality of the spatio-temporal matrices. DMD can tackle this problem by using the singular value decomposition (SVD) of \(H_{1}\) \cite{kutz2016dynamic}.

By taking SVD we can write \(
H_1 \;=\; U \,\Sigma\, V^*\). Here,
\(U\in\mathbb{R}^{mn\times q},\; \Sigma\in\mathbb{R}^{q\times q},\; V\in\mathbb{R}^{q\times mn},
\). Consequently,
\(
H_1^{\dagger} \;=\; ( U \,\Sigma\, V^*)^{\dagger}\; =\; V\,\Sigma^{-1}\,U^*.
\)
Using the relation in Equation~\ref{eq:Ah} a least-squares estimate of \(A\) can be written as:
\[
A \approx H_2 H_1^{\dagger} \;=\; H_2 V \Sigma^{-1} U^*.
\]
We can further reduce the dimensionality by truncating smaller ranks and keeping the \(r\) largest singular values. Projecting \(A\) onto the \(r\)-dimensional subspace spanned by the columns of \(U\) yields the reduced operator
\[
\tilde{A} \;=\; U_{r}^* A U_{r} \;=\; U_{r}^* H_2 V_{r} \Sigma^{-1} \in \mathbb{R}^{r\times r}.
\]

We compute the eigen decomposition of the reduced operator:
\[
\tilde{A} W \;=\; W \Lambda, \qquad
\Lambda=\operatorname{diag}(\lambda_1,\dots,\lambda_r),\; W\in\mathbb{C}^{r\times r}
\]
The DMD modes in the full state space are obtained as
\[
\Phi \;=\; H_2 V \Sigma^{-1} W \in \mathbb{C}^{\,mn\times r}
\]
so that the \(k\)-th DMD mode \(\phi_k\) is the \(k\)-th column of \(\Phi\) and can be un-vectorized to the spatial field
\(\phi_k(x,y)=\operatorname{vec}^{-1}(\phi_k)\in\mathbb{C}^{m\times n}.\) One can convert discrete eigenvalues \((\lambda_{k})\) to continuous-time eigenvalues (\((\omega_{k})\)) as follows:

\begin{equation} \label{eq:period}
    \omega_k \;=\; \frac{\ln(\lambda_k)}{\Delta t}
\end{equation}

where \(\Delta t\) is the sampling interval. We collect all the eigenvalues into \(\Omega=\operatorname{diag}(\omega_1,\dots,\omega_r)\). We define the modal amplitude vector \(\alpha\in\mathbb{C}^r\) from the initial snapshot. If no mean-removal is used, a common choice is
\[
\alpha \;=\; \Phi^{\dagger}\, h_1,
\]
where \(\Phi^{\dagger}\) is the Moore--Penrose pseudo-inverse of \(\Phi\). Then the reconstructed vectorized state at continuous time \(t\) is
\[
\hat{h}(t) \;=\; \Phi\,\big(e^{\Omega t}\alpha\big)
\qquad\text{(continuous-time form)}
\]
or, in discrete-time at snapshot index \(j\) (with \(t_j=(j-1)\Delta t\)),
\[
\hat{h}_j \;=\; \Phi\,\Lambda^{\,j-1}\alpha
\qquad\text{(discrete-time form).}
\]
Finally, we reshape the vector back to the spatial field:
\[
\widehat{\eta}(x,y,t) \;=\; \operatorname{vec}^{-1}\!\big(\hat{h}(t)\big) \in \mathbb{R}^{m\times n}.
\]
\begin{figure}
    \centering
    \includegraphics[width=1.0\linewidth]{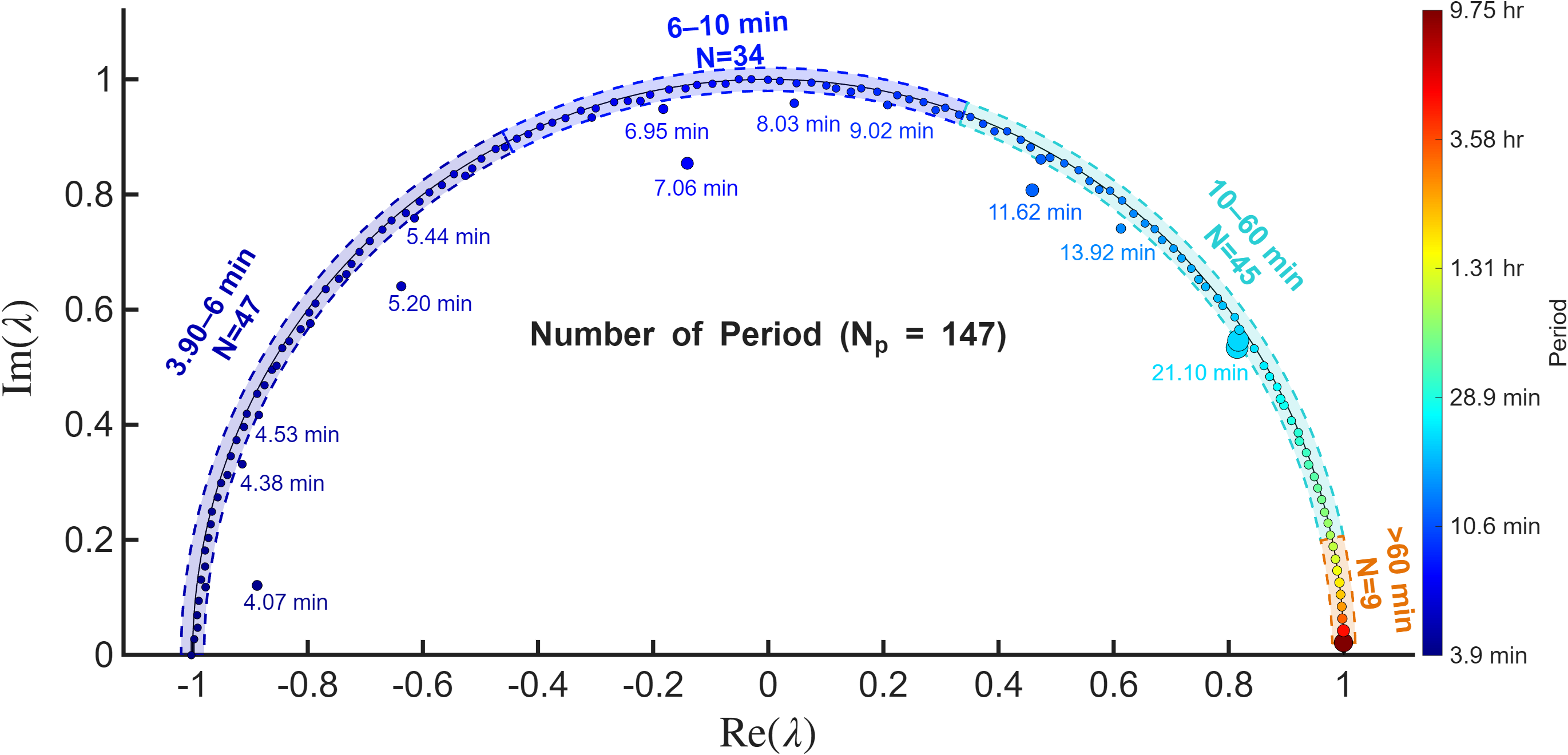}
    \caption{Semicircular distribution of DMD eigenspectra estimated from spatio-temporal bed elevation data. Due to symmetry, only the upper half of the unit circle is shown. The semicircle is partitioned into four regions based on period ranges. Here, \(N\) denotes the number of eigenvalues within each region. Eigenvalues inside the unit semicircle are annotated with their individual periods, while those on the unit semicircle are color-coded according to period.}
    \label{fig:eigenvalue}
\end{figure}

It is a common practice in DMD analysis to remove the temporal mean from the data to focus on the fluctuations. Removing the mean elevation we get,
\[
\bar{\eta} \;=\; \frac{1}{K}\sum_{k=1}^K \eta_k(x,y), \qquad
\bar{h} \;=\; \operatorname{vec}(\bar{\eta}) \in\mathbb{R}^{mn},
\]
and form fluctuation snapshots
\[
\tilde{h}_k \;=\; h_k - \bar{h}, \qquad
\tilde{H}_1=[\tilde{h}_1,\dots,\tilde{h}_{K-1}],\;
\tilde{H}_2=[\tilde{h}_2,\dots,\tilde{h}_K].
\]
Apply the SVD and DMD steps above to \(\tilde{H}_1,\tilde{H}_2\) to obtain fluctuation modes \(\tilde{\Phi}\), eigenvalues \(\tilde{\Lambda}\), and amplitudes \(\tilde{\alpha}\) (computed for example by \(\tilde{\alpha}=\tilde{\Phi}^{\dagger}\tilde{h}_1\)). The full-field reconstruction then adds the mean back:
\[
\widehat{h}(t) \;=\; \bar{h} \;+\; \tilde{\Phi}\,e^{\Omega t}\,\tilde{\alpha},
\]
or equivalently at discrete snapshot \(j\),
\[
\widehat{h}_j \;=\; \bar{h} \;+\; \tilde{\Phi}\,\tilde{\Lambda}^{\,j-1}\tilde{\alpha}.
\]
Finally,
\begin{equation} \label{eq:reconstruction}
      \widehat{\eta}(x,y,t_j) \;=\; \operatorname{vec}^{-1}\!\big(\widehat{h}_j\big).    
\end{equation}

Here, $\phi_k(x,y)$ encodes coherent spatial patterns, $\omega_k$ determines their temporal evolution, and $\alpha_k$ sets their relative amplitudes.

\subsection{DMD Results}
We have applied DMD on the bed elevation dynamics described in Section~\ref{section:data}. The dataset has \(300\) snapshots. We have selected \(98\%\) of the data and took full rank in SVD. It has been observed that the elevation dynamics are sensitive to rank reduction. Discarding ranks can significantly distort the eigenspectrum by introducing spurious values. Since DMD attempts to preserve the Frobenius norm of the \(A\) matrix, it can introduce spurious eigenvalues. This significantly deteriorates the reconstruction performance. At the same time, it distorts the eigenspectra, which impairs accurate system identification. Therefore, we selected full rank. In the full rank SVD, the maximum number of possible eigenvalue is \(294\) (\(98\%\) of \(300\) snapshots). Since each eigenvalue has a complex conjugate, we obtained \(147\) different frequencies (time period, \(T=1/f\)) from the analysis. 

\begin{figure}
    \centering
    \includegraphics[width=1.0\linewidth]{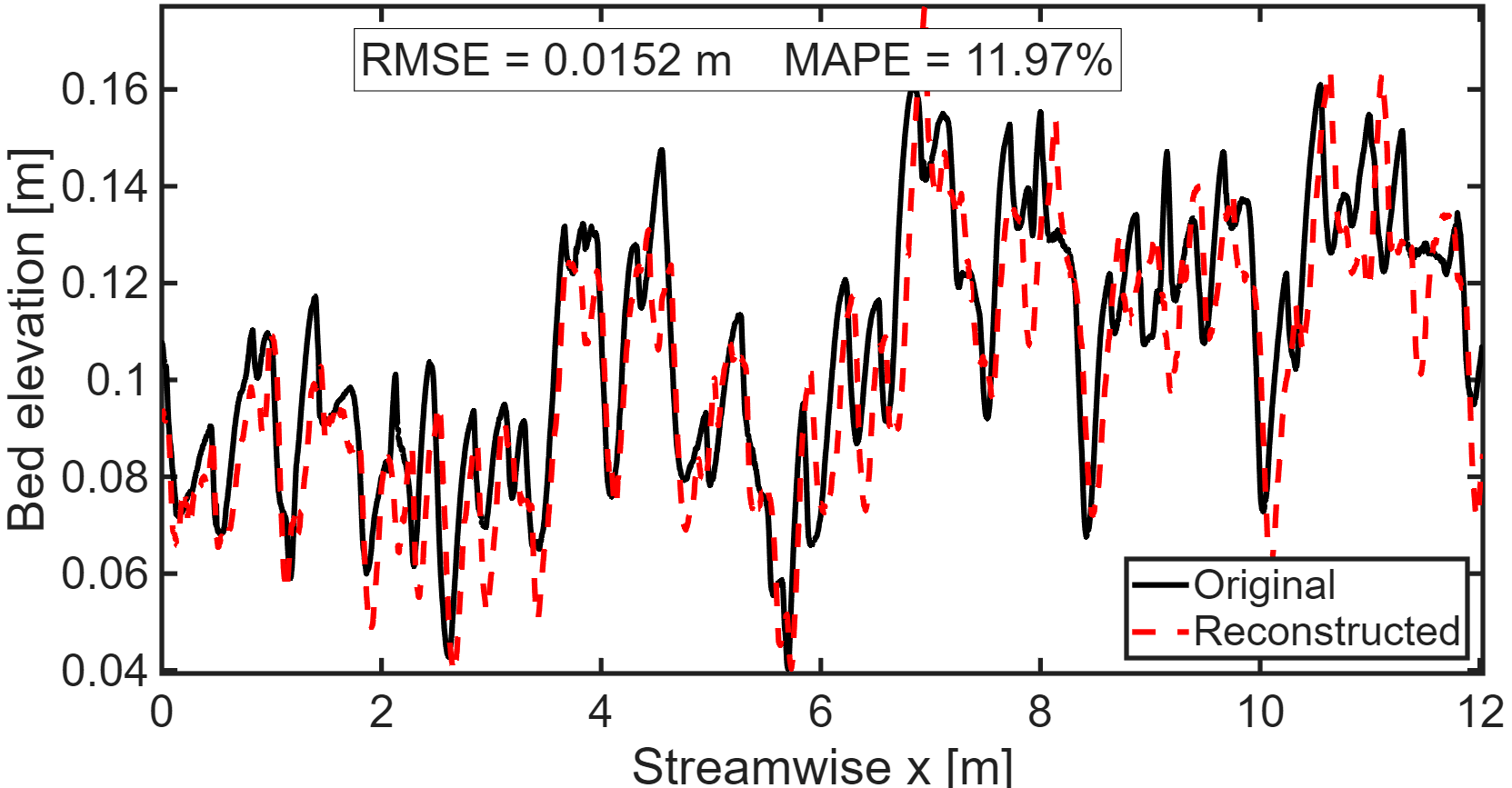}
    \caption{Comparison of real river bed elevation and that of DMD reconstructed snapshot at a transect.}
    \label{fig:reconstruction}
\end{figure}

\begin{figure}
    \centering
    \includegraphics[width=1.0\linewidth]{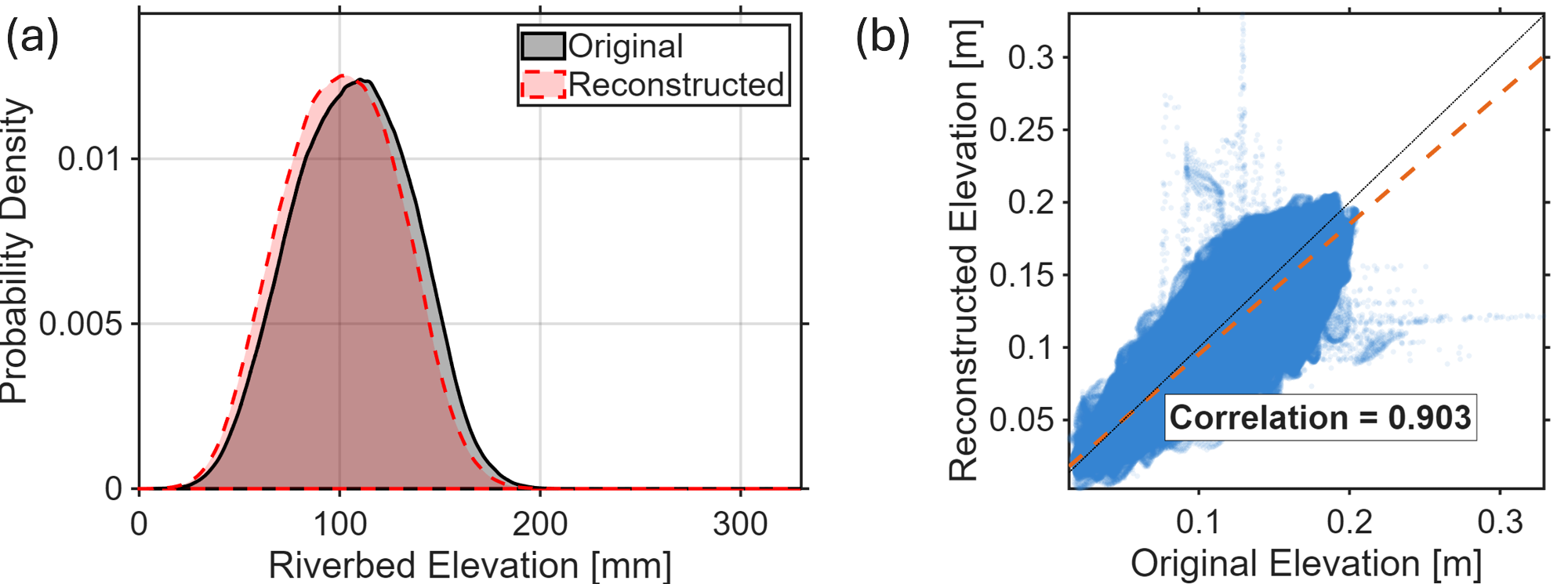}
    \caption{(a) Comparison of the probability distribution function between the real river bed elevation and that of the DMD reconstructed. (b) Scatter plot correlation between the original riverbed elevation and that of the DMD reconstructed.}
    \label{fig:PDF_Scatter}
\end{figure}

\begin{figure}
    \centering
    \includegraphics[width=1.0\linewidth]{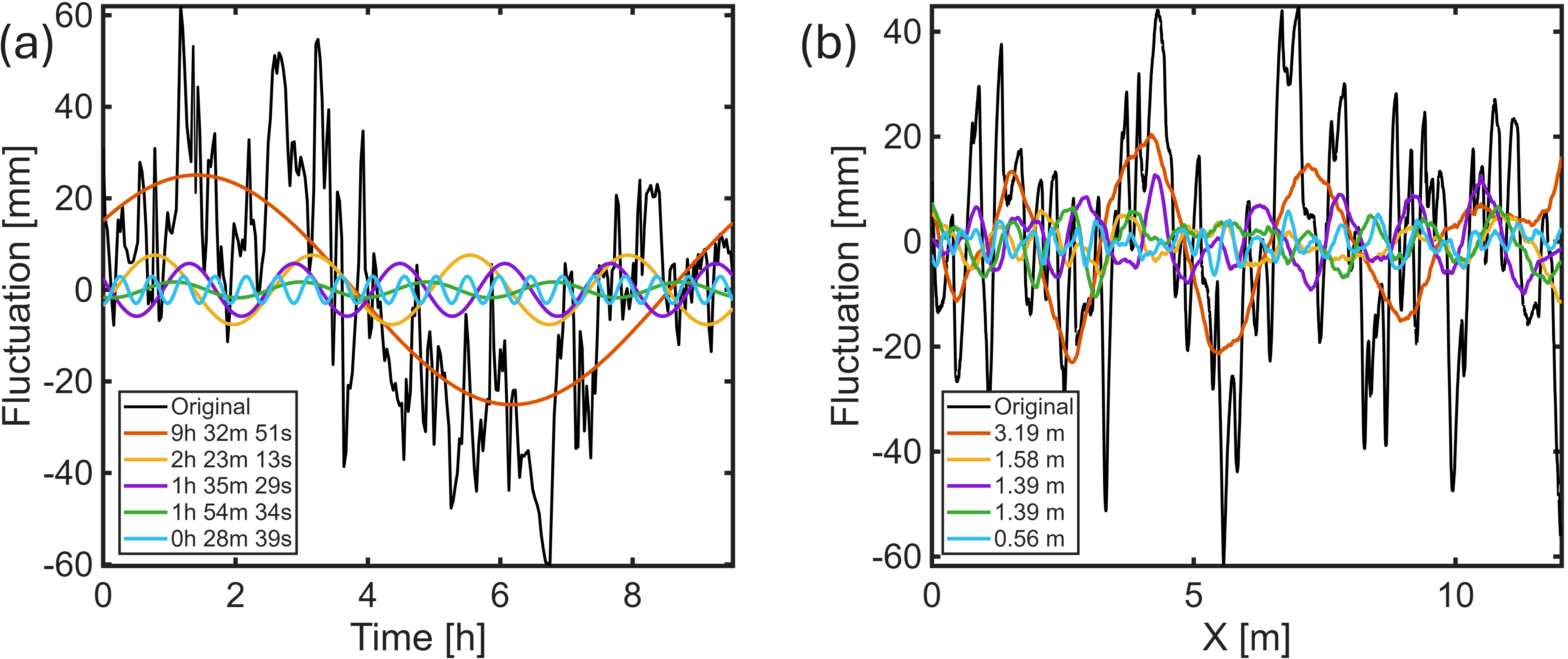}
    \caption{(a) Elevation fluctuation and dominant top five DMD modes that contribute most to the sediment transportation are shown at a fixed point over time. (b) Elevation fluctuation and the dominant top five DMD modes that contribute most to sediment transportation are shown at a fixed instant across the mid-transect.}
    \label{fig:mode_point}
\end{figure}

Figure~\ref{fig:eigenvalue} shows the DMD spectra. It also shows the periods associated with identified DMD eigenvalues. The periods were estimated using Equation~\ref{eq:period}. The size of the dots representing the eigenvalue is proportional to the DMD power associated with the mode. This helps identify the strength of the mode and its temporal period. In general, DMD eigenvalues are plotted on a unit circle. Since DMD eigenvalues exhibit symmetric behavior along x-axis, for brevity we plot only the upper half. We divide the semicircle into four regions based on periods. From left to right, along the semicircle, frequency decreases and time period increases. The divided regions are shown using separate envelopes. The figure illustrates the range of time periods in each envelope and the number of eigenvalues that fall inside the envelope. Here, if any eigenvalue (\(\lambda\)) varies by less than \(1\%\), i.e., \(0.99 < |\lambda| < 1.01\), we consider it to lie on the unit circle. Such an eigenvalue is associated with neither growth nor decay. In other words, it is considered persistent. On the other hand, eigenvalues that lie within the unit circle are shown with periods associated with them. These eigenvalues have relatively larger DMD power. Most of these eigenvalues are associated with modes shorter than \(10\) minutes, indicating a fast-migrating and temporally decaying component of the bedform. In total, there are \(12\) eigenvalues inside the unit circle, while \(135\) eigenvalues lie on or outside the unit circle. The range \(3.90-6\; \text{min}\) contains the majority of the periods (\(N=47\)), while only \(9\) eigenvalues have periods longer than \(1 \; \text{hour}\). This indicates that bedform migration dynamics are primarily dominated by shorter periods or high-frequency components. To avoid clutter, we did not annotate individual periods on and outside the unit circle. Instead, the periods are represented using a color scale.
  
Figure~\ref{fig:reconstruction} shows the reconstructed bed elevation at \(y=63\) for the 150th time snapshot. The MAPE (mean absolute percentage error) estimated for that snapshot is \(11.97 \%\), showing the reconstruction accuracy of DMD. The figure shows that DMD is capable of visually reconstructing the bed elevation. The reconstruction was done using Equation~\ref{eq:reconstruction}. To examine the overall reconstruction performance, we estimated statistical measures of the reconstructed dynamics by computing the probability distribution function (PDF) and scatter plot. Figure~\ref{fig:PDF_Scatter}(a) compares the original PDF and reconstructed PDF. It shows that the reconstructed PDF almost overlaps the original. Figure~\ref{fig:PDF_Scatter}(b) shows the correlation between the original and the reconstructed elevation dynamics. Each point in the correlation signifies one point \(\eta_{k}(x,t)\) of the bed elevation field at a particular instant. The correlation score is \(0.9\), indicating the identified eigenvalues can accurately reconstruct the original dynamics. This is important in the sense that it ensures the system identification is justifiable.

Figure~\ref{fig:mode_point}(a) shows the evolution of selected DMD modes at a particular point. The point is located at the center of the channel. The figure shows temporal evolution of elevation at that point. The black line represents the fluctuation of original elevation, while the colored lines show the temporal evolution of DMD modes associated with the point. Note that, in DMD, the mean is subtracted from the data. Therefore, DMD modes are computed from the fluctuations and compared with fluctuations rather than the raw elevation.  Figure~\ref{fig:mode_point}(b) characterizes the spatial behavior of the selected modes. Here we show the spatial component of the modes along one transect (\(y=63\)) at \(150\)th snapshot. The figure also illustrates the wavelengths associated with them. Note that the legend in Figure~\ref{fig:mode_point} (a) and Figure~\ref{fig:mode_point} (b) follows the same order.

% \begin{figure}
%     \centering
%     \includegraphics[width=1.0\linewidth]{midtransect_top5_Cnet_modes_LS.png}
%     \caption{Elevation fluctuation and dominant top five DMD modes that contribute most to the sediment transportation are shown at a fixed instant across the mid-transect.}
%     \label{fig:mode_instant}
% \end{figure}

\section{Linking DMD and Sediment Flux via the Exner Equation} \label{section:link}

We begin with the two-dimensional Exner equation, which represents mass balance in sediment \cite{paola2005generalized,jerolmack2005unified,mcelroy2009nature}:
\begin{equation}
(1-\lambda_p)\,\frac{\partial \eta}{\partial t} \;+\; \nabla\!\cdot\vec q_s \;=\; 0
\label{eq:exner}
\end{equation}
where $\eta(x,y,t)$ is the bed elevation, $\lambda_p$ is the bed porosity, and $\vec q_s=(q_{s,x},q_{s,y})$ is the volumetric sediment flux vector. From DMD analysis in Section~\ref{sec:math}, we obtain a linear, data-driven approximation of the temporal evolution of the topography. In continuous-time form we write
\begin{equation}
\frac{\partial\eta}{\partial t} \;\approx\; L\eta
\label{eq:dmd_ct}
\end{equation}
where $L$ denotes the (linear) generator associated with the DMD model. In discrete time, DMD yields a matrix $A$ with eigenpairs $A\phi_k=\lambda_k\phi_k$; the corresponding continuous-time generator is related by $L=\tfrac{1}{\Delta t}\log(A)$, so that $L\phi_k=\omega_k\phi_k$ with $\omega_k=\log(\lambda_k)/\Delta t$.
 Substituting \eqref{eq:dmd_ct} into \eqref{eq:exner} yields
\begin{equation}\label{eq:fundamental}
(1-\lambda_p)\,L\eta \;+\; \nabla\!\cdot\vec q_s \;=\; 0 \implies
\nabla\!\cdot\vec q_s \;=\; - (1-\lambda_p)\,L\eta.
\end{equation}

To relate the DMD representation of $\eta$ to net streamwise flux through a cross-section at a fixed $y$, we integrate \eqref{eq:fundamental} in $x$ from $x_{\min}$ to $x_{\max}$:
\begin{equation}
\int_{x_{\min}}^{x_{\max}} \nabla\!\cdot\vec q_s(x,y,t)\,dx
\;=\; - (1-\lambda_p)\int_{x_{\min}}^{x_{\max}} (L\eta)(x,y,t)\,dx.
\label{eq:integral}
\end{equation}

The left-hand side of the Exner equation can be written using the 2D divergence as:  
\begin{align}
&\int_{x_{\min}}^{x_{\max}} \nabla\!\cdot\vec q_s(x,y,t)\,dx
   = \int_{x_{\min}}^{x_{\max}}
     \left(\frac{\partial q_{s,x}}{\partial x} 
           + \frac{\partial q_{s,y}}{\partial y}\right) dx \notag \\
&= \frac{\partial}{\partial x}
     \left(\int_{x_{\min}}^{x_{\max}} q_{s,x}\,dx\right)
   + \frac{\partial}{\partial y}
     \left(\int_{x_{\min}}^{x_{\max}} q_{s,y}\,dx\right) \notag \\
&= q_{s,x}^{\mathrm{net}}(y,t)
   + \frac{\partial}{\partial y}
     \left(\int_{x_{\min}}^{x_{\max}} q_{s,y}\,dx\right).
\label{eq:LHS_identity}
\end{align}

We have defined the net streamwise flux through the transect as:  
\[
q_{s,x}^{\mathrm{net}}(y,t) := q_{s,x}(x_{\max},y,t)-q_{s,x}(x_{\min},y,t)
\]  
We obtain from Equation~\ref{eq:LHS_identity} and Equation~\ref{eq:integral}  
\begin{equation}
q_{s,x}^{\mathrm{net}}(y,t) + 
\frac{\partial}{\partial y}\!\left(\int_{x_{\min}}^{x_{\max}} q_{s,y}\,dx\right)
= - (1-\lambda_p)\int_{x_{\min}}^{x_{\max}} (L\eta)\,dx .
\label{eq:exact_relation}
\end{equation}

Since streamwise transport dominates and lateral flux variations are negligible, the \(\partial_y\) term is commonly dropped. This yields the leading-order relation  
\begin{equation}
q_{s,x}^{\mathrm{net}}(y,t) \;=\; - (1-\lambda_p)\int_{x_{\min}}^{x_{\max}} (L\eta)\,dx .
\label{eq:exact_relation_dmd}
\end{equation}

% \begin{table}[ht]
% \centering
% \caption{Summary of variables and notation}
% \begin{tabular}{ll}
% \hline
% Symbol & Description \\
% \hline
% $\eta(x,y,t)$     & Bed elevation \\
% $\lambda_p$       & Bed porosity (a constant) \\
% $\vec q_s=(q_{s,x},q_{s,y})$ & Sediment flux  \\
% $q_{s,x}^{\mathrm{net}}(y,t)$ & Net streamwise sediment flux through a transect \\
% $L$               & Linear generator (continuous-time DMD operator) \\
% $\phi_k(x,y)$     & Spatial DMD mode \\
% $\alpha_k$        & Initial modal amplitude  \\
% $\omega_k$       & Continuous-time growth/decay rate  \\
% $\lambda_k$       & Discrete-time DMD eigenvalue \\
% $\Delta t$        & Sampling interval \\
% $x_{\min},x_{\max}$ & Streamwise integration bounds \\
% $y$               & Cross-stream coordinate \\
% $t$               & Time \\
% \hline
% \end{tabular}
% \label{tab:variables}
% \end{table}

We can express the reconstructed bed elevation as a superposition of DMD modes,
\[
\eta(x,y,t)\approx\sum_k \alpha_k \phi_k(x,y)\,e^{\omega_k t}
\]
where $\{\phi_k\}$ are (continuous-time) DMD modes, $\alpha_k$ are initial amplitudes, and $\omega_k$ are the associated continuous growth rates (units $[T]^{-1}$). Under the assumptions that (i) $L$ is linear and (ii) commuting $L$ and the $x$-integral is permissible, the right-hand side of \eqref{eq:exact_relation} becomes

\begin{align}
q_{s,x}^{\mathrm{net}}(y,t) &\approx- (1-\lambda_p)\!\int_{x_{\min}}^{x_{\max}} (L\eta)\,dx\\
&\approx - (1-\lambda_p)\!
   \sum_k \alpha_k e^{\omega_k t}
   \int_{x_{\min}}^{x_{\max}} L\phi_k(x,y)\,dx \nonumber\\
&\approx - (1-\lambda_p)\!
   \sum_k \alpha_k e^{\omega_k t}\,\omega_k
   \int_{x_{\min}}^{x_{\max}} \phi_k(x,y)\,dx
\label{eq:RHS_modes}
\end{align}

Here, we used the eigenvalue relation $L\phi_k = \omega_k \phi_k$. Equation \eqref{eq:RHS_modes} shows that a weighted sum over DMD modes gives the net streamwise sediment flux through a transect. Each mode contributes in proportion to its integrated spatial mode shape over the transect, its modal amplitude, and its growth/decay rate $\omega_k$. We have assumed that cross-stream transport and its lateral gradient are negligible at the transect of interest. 
% By combining \eqref{eq:exact_relation} and \eqref{eq:exact_relation_dmd}, we obtain:

% \begin{equation}
% q_{s,x}^{\mathrm{net}}(y,t)
% \approx - (1-\lambda_p)\sum_k \alpha_k e^{\omega_k t}\,\omega_k
%     \left(\int_{x_{\min}}^{x_{\max}} \phi_k(x,y)\,dx\right)
% \label{eq:approx_final}
% \end{equation}

\begin{figure}[tb]
    \centering
    \includegraphics[width=1.0\linewidth]{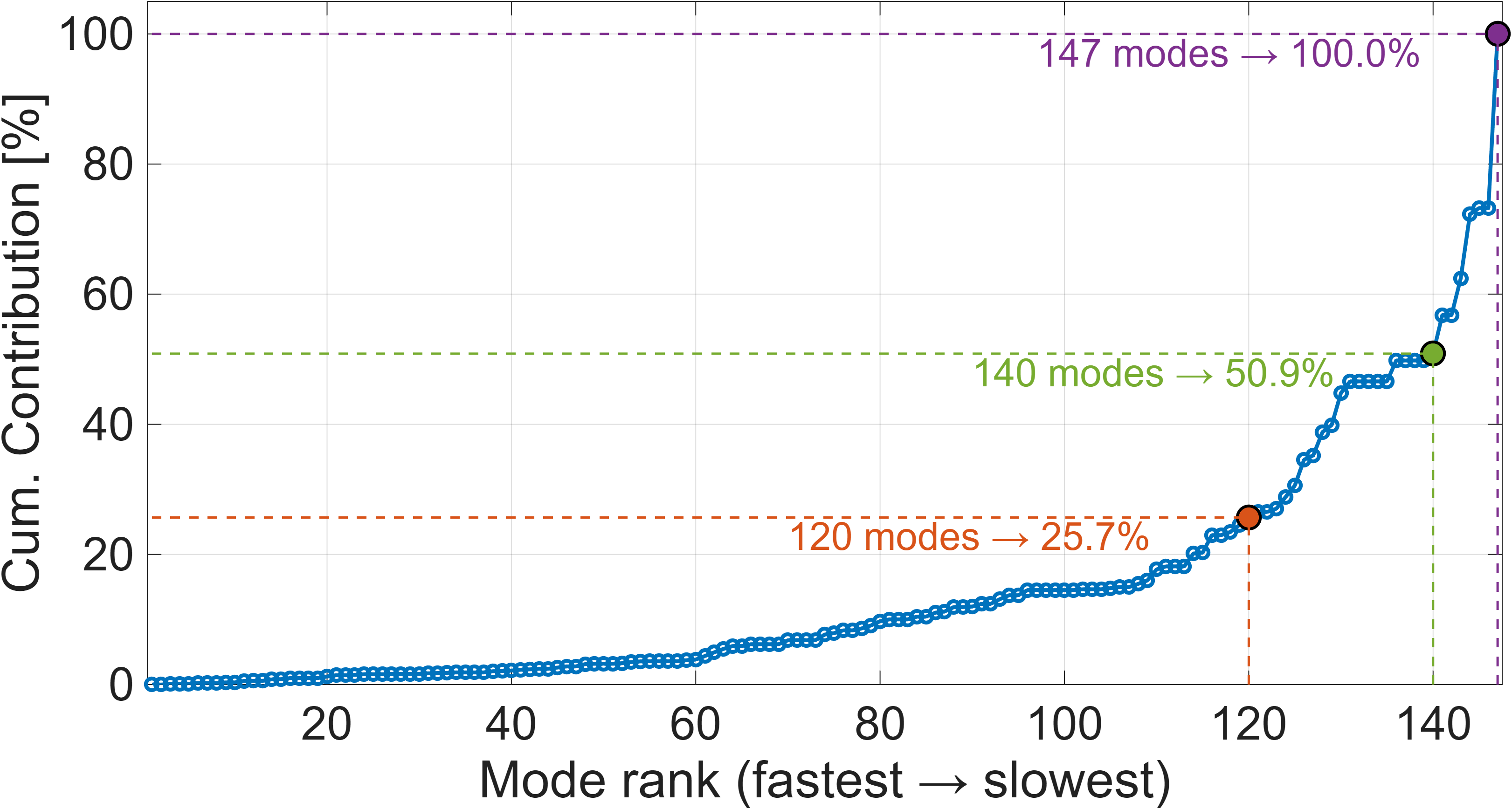}
    \caption{%(a) Correlation between DMD power and \(C_{net}\), showing strong positive dependence (\(\rho = 0.867\)); colors denote modal wavelength \(\lambda\),(b) Relationship between wavelength and \(C_{net}\), indicating positive correlation (\(\rho = 0.723\)); color represents modal frequency \(f\), (c) Correlation between modal speed and \(C_{net}\), showing negative dependence (\(\rho = -0.412\)); colors denote wavelength \(\lambda\),
    Cumulative contribution of individual modes to total sediment transportation; the slowest 7 modes account for 50.9\%, the fastest 120 modes contribute 25.7\%, and all 147 modes explain 100\% of transport.}
    \label{fig:evalutation}
\end{figure}

\section{Discussion and Implications}
In this section, we evaluate the implications of the mathematical derivation in Equation~\ref{eq:RHS_modes}. The goal is to verify whether the surrogate measure of sediment flux estimated from DMD modes is consistent with the literature. By discarding the summation sign in Equation~\ref{eq:RHS_modes}, we can write the contribution of each DMD mode \((\phi_{k})\) to the net sediment transport as \(q_{s,x}^{k}(y,t,\phi_{k}) = - (1-\lambda_p) \alpha_k e^{\omega_k t}\,\omega_k \left(\int_{x_{\min}}^{x_{\max}} \phi_k(x,y)\,dx\right).\) Here, \(q^{k}_{s,x}(y,t,\phi_{k})\) signifies the individual modal contribution to the net sediment flux, allowing the net flux to be decomposed into multiple spatial and temporal scales. The speed associated with each mode at a transect was estimated by computing its wavelength using spatial Fourier and Hilbert analysis, then multiplying it by the frequency obtained from the DMD eigenvalues.

The modes shown in Figure~\ref{fig:mode_point} were selected based on their contribution to sediment transport. The figure shows that slower modes contribute most individually, as \(4\) of the \(5\) modes depicted have a period longer than \(1 \; \text{hour}\). We sort the modes based on their speed and compute cumulative contributions by summing the sorted percentile contributions from the fastest to the slowest mode. This provides a clear visualization of how a small subset of modes dominates total sediment transport.

Figure~\ref{fig:evalutation} presents the cumulative contribution
of individual modes to net sediment transport. The analysis
shows that just the \(7\) slowest modes account for \(50.9\%\)
of the total transport, while the remaining \(140\) modes
collectively contribute the other half. Moreover, the fastest
\(120\) modes, although numerous, contribute only \(25.7\%\)
in total. This indicates that sediment transport is dominated
by a small number of slow, long-wavelength modes, whereas
faster, short-wavelength modes provide a distributed but
comparatively smaller contribution. Such behavior is
consistent with recent studies showing that the spectral
contribution of bedforms to sediment flux decreases with
increasing frequency following a power-law scaling
\(q_s(f)\propto f^{-0.6}\) \cite{lee2023scaling,lee2022reconstructing}. Because
low-frequency components correspond to large-scale,
slowly migrating bedforms, this relationship implies that
a small number of slow modes can carry a disproportionately
large fraction of the sediment transport. In contrast,
higher-frequency modes associated with smaller bedforms
contribute more modestly despite their larger number.
Unlike purely spectral approaches, the present DMD-based
framework directly links sediment transport to dynamically
evolving spatio-temporal modes of riverbed evolution,
allowing the contribution of each mode to the net transport
to be quantified explicitly.

\section{Conclusion}
In this paper, we present a novel data-driven framework for characterizing spatiotemporal mode-dependent sediment transport from riverbed topography. By substituting DMD modes of bed elevation into the Exner equation, we establish a physics-informed link between spatio-temporal modes and net sediment flux. Our analysis shows that the largest contributions to sediment transport come from slow, long-wavelength modes, while faster, small-wavelength modes, although individually smaller, are numerous and collectively influence transport dynamics. The analysis further reveals that persistent modes, characterized by DMD eigenvalues on the unit circle, act as the primary carriers of the bulk of the sediment flux. Thus, the proposed framework provides a non-intrusive surrogate for measuring sediment flux, offering new insights into the multiscale structure of bedform-driven transport. Future work will focus on extending the methodology to characterize sediment transport under different flow and channel conditions. 

\bibliographystyle{IEEEtran}
\bibliography{ACC_ref}

@article{lee2022reconstructing,
  title={Reconstructing sediment transport by migrating bedforms in the physical and spectral domains},
  author={Lee, Jiyong and Singh, Arvind and Guala, Michele},
  journal={Water Resources Research},
  volume={58},
  number={7},
  pages={e2022WR031934},
  year={2022},
  publisher={Wiley Online Library}
}

@article{mustavee2025koopman,
  title={A Koopman-Theoretic Approach to Car-Following and Multi-Lane Interaction Modeling},
  author={Mustavee, Shakib and Agarwal, Shaurya},
  journal={IEEE Open Journal of Intelligent Transportation Systems},
  year={2025},
  publisher={IEEE}
}

@article{mustavee2022linear,
  title={A linear dynamical perspective on epidemiology: interplay between early COVID-19 outbreak and human mobility},
  author={Mustavee, Shakib and Agarwal, Shaurya and Enyioha, Chinwendu and Das, Suddhasattwa},
  journal={Nonlinear Dynamics},
  volume={109},
  number={2},
  pages={1233--1252},
  year={2022},
  publisher={Springer}
}

@article{lee2023scaling,
  title={On the scaling and growth limit of fluvial dunes},
  author={Lee, Jiyong and Singh, Arvind and Guala, Michele},
  journal={Journal of Geophysical Research: Earth Surface},
  volume={128},
  number={6},
  pages={e2022JF006955},
  year={2023},
  publisher={Wiley Online Library}
}

@article{higham2018implications,
  title={Implications of the selection of a particular modal decomposition technique for the analysis of shallow flows},
  author={Higham, JE and Brevis, W and Keylock, CJ},
  journal={Journal of Hydraulic Research},
  volume={56},
  number={6},
  pages={796--805},
  year={2018},
  publisher={Taylor \& Francis}
}

@article{vericat2006bed,
  title={Bed load bias: Comparison of measurements obtained using two (76 and 152 mm) Helley-Smith samplers in a gravel bed river},
  author={Vericat, Damia and Church, Michael and Batalla, Ramon J},
  journal={Water Resources Research},
  volume={42},
  number={1},
  year={2006},
  publisher={Wiley Online Library}
}

@book{helley1971development,
  title={Development and calibration of a pressure-difference bedload sampler},
  author={Helley, Edward J and Smith, Winchell},
  year={1971},
  publisher={US Department of the Interior, Geological Survey, Water Resources Division}
}

@article{wang2016reduced,
  title={Reduced sediment transport in the Yellow River due to anthropogenic changes},
  author={Wang, Shuai and Fu, Bojie and Piao, Shilong and L{\"u}, Yihe and Ciais, Philippe and Feng, Xiaoming and Wang, Yafeng},
  journal={Nature Geoscience},
  volume={9},
  number={1},
  pages={38--41},
  year={2016},
  publisher={Nature Publishing Group UK London}
}

@article{singh2011multiscale,
  title={Multiscale statistical characterization of migrating bed forms in gravel and sand bed rivers},
  author={Singh, Arvind and Lanzoni, Stefano and Wilcock, Peter R and Foufoula-Georgiou, Efi},
  journal={Water Resources Research},
  volume={47},
  number={12},
  year={2011},
  publisher={Wiley Online Library}
}

@book{kutz2016dynamic,
  title={Dynamic mode decomposition: data-driven modeling of complex systems},
  author={Kutz, J Nathan and Brunton, Steven L and Brunton, Bingni W and Proctor, Joshua L},
  year={2016},
  publisher={SIAM}
}

@article{paola2005generalized,
  title={A generalized Exner equation for sediment mass balance},
  author={Paola, Chris and Voller, Vaughan R},
  journal={Journal of Geophysical Research: Earth Surface},
  volume={110},
  number={F4},
  year={2005},
  publisher={Wiley Online Library}
}

@article{jerolmack2005unified,
  title={A unified model for subaqueous bed form dynamics},
  author={Jerolmack, Douglas J and Mohrig, David},
  journal={Water Resources Research},
  volume={41},
  number={12},
  year={2005},
  publisher={Wiley Online Library}
}

@article{sharma2016correspondence,
  title={Correspondence between Koopman mode decomposition, resolvent mode decomposition, and invariant solutions of the Navier-Stokes equations},
  author={Sharma, Ati S and Mezi{\'c}, Igor and McKeon, Beverley J},
  journal={Physical Review Fluids},
  volume={1},
  number={3},
  pages={032402},
  year={2016},
  publisher={APS}
}

@article{singh2009experimental,
  title={Experimental evidence for statistical scaling and intermittency in sediment transport rates},
  author={Singh, Arvind and Fienberg, Kurt and Jerolmack, Douglas J and Marr, Jeffrey and Foufoula-Georgiou, Efi},
  journal={Journal of Geophysical Research: Earth Surface},
  volume={114},
  number={F1},
  year={2009},
  publisher={Wiley Online Library}
}

@article{singh2012coupled,
  title={Coupled dynamics of the co-evolution of gravel bed topography, flow turbulence and sediment transport in an experimental channel},
  author={Singh, Arvind and Foufoula-Georgiou, Efi and Port{\'e}-Agel, Fernando and Wilcock, Peter R},
  journal={Journal of Geophysical Research: Earth Surface},
  volume={117},
  number={F4},
  year={2012},
  publisher={Wiley Online Library}
}

@article{guala2020mixed,
  title={A mixed length scale model for migrating fluvial bedforms},
  author={Guala, Michele and Heisel, Michael and Singh, Arvind and Musa, Mirko and Buscombe, Daniel and Grams, Paul},
  journal={Geophysical Research Letters},
  volume={47},
  number={15},
  pages={e10--1029},
  year={2020},
  publisher={Wiley Online Library}
}

@article{guala2014spectral,
  title={Spectral description of migrating bed forms and sediment transport},
  author={Guala, Michele and Singh, Arvind and BadHeartBull, Nicholas and Foufoula--Georgiou, Efi},
  journal={Journal of Geophysical Research: Earth Surface},
  volume={119},
  number={2},
  pages={123--137},
  year={2014},
  publisher={Wiley Online Library}
}

@article{mcelroy2009nature,
  title={Nature of deformation of sandy bed forms},
  author={McElroy, Brandon and Mohrig, David},
  journal={Journal of Geophysical Research: Earth Surface},
  volume={114},
  number={F3},
  year={2009},
  publisher={Wiley Online Library}
}

@article{bunte2005effect,
  title={Effect of sampling time on measured gravel bed load transport rates in a coarse-bedded stream},
  author={Bunte, Kristin and Abt, Steven R},
  journal={Water Resources Research},
  volume={41},
  number={11},
  year={2005},
  publisher={Wiley Online Library}
}

@book{simons1965bedload,
  title={Bedload equation for ripples and dunes},
  author={Simons, Daryl B and Richardson, Everett V and Nordin, Carl F},
  year={1965},
  publisher={US Government Printing Office}
}

@article{ranjbar2020entropy,
  title={Entropy and intermittency of river bed elevation fluctuations},
  author={Ranjbar, Sevil and Singh, Arvind},
  journal={Journal of Geophysical Research: Earth Surface},
  volume={125},
  number={8},
  pages={e2019JF005499},
  year={2020},
  publisher={Wiley Online Library}
}

\end{document}